\documentclass[prb,twocolumn,superscriptaddress,showpacs]{revtex4}
\usepackage{amsmath,amssymb,graphicx,epsfig,color, float}
\usepackage{epstopdf}

\begin{document}
\title{Theory of an excitonic THz laser with two-photon excitation}

\author{A. A. Pervishko}
\affiliation{Division of Physics and Applied Physics, Nanyang Technological University 637371, Singapore}

\author{T. C. H. Liew}
\affiliation{Division of Physics and Applied Physics, Nanyang Technological University 637371, Singapore}
\affiliation{Mediterranean Institute of Fundamental Physics, 31 via Appia Nuova, 00040 Rome, Italy}

\author{A. V. Kavokin}
\affiliation{Spin Optics Laboratory, Saint-Petersburg State University, Petrodvorets, 198504 St. Petersburg, Russia}
\affiliation{Physics and Astronomy School, University of Southampton, Highfield, Southampton SO17 IBJ, UK}

\author{I. A. Shelykh}
\affiliation{Division of Physics and Applied Physics, Nanyang Technological University 637371, Singapore}
\affiliation{Science Institute, University of Iceland, Dunhagi-3, IS-107, Reykjavik, Iceland}

\date{\today}

\begin{abstract}
We consider theoretically THz lasing in a system consisting of a quantum well placed inside an optical microcavity and a THz cavity in the regime of two-photon excitation of 2p dark exciton states. The stability of the system with varying parameters of the microcavity under coherent pumping is studied. We show that the nonlinearity provided by two photon absorption can give rise to bistability and hysteresis in the THz output.
\end{abstract}

\pacs{78.67.Pt,78.66.Fd,78.45.+h}

\maketitle

\section{Introduction}

The possibility to design efficient sources of terahertz (THz) radiation attracts the attention of researches working in various scientific fields. This is connected with the wide possibilities of using the THz range of the electromagnetic spectrum, which is not limited to purely scientific problems. The current use of THz radiation includes fields such as medicine, security, biosensing, and others~\cite{Dragoman,Mueller}.

However, none of the currently existing THz emitters universally satisfies the application
requirements. For example, emitters based on nonlinear-optical frequency
down-conversion are bulky, expensive, and power consuming. Various semiconductor
\cite{Hu} and carbon-based \cite{Portnoi,Wright} devices based upon
intraband optical transitions are compact, but have a low quantum efficiency. The crucial factor that
restricts the efficiency of semiconductor THz sources is the large characteristic time of spontaneous emission of THz photons (typically milliseconds) compared to the much shorter lifetime of the involved electronic states (typically fractions of a nanosecond). Attempted ways to improve the situation include the use of the Purcell effect\cite{Purcell,Gerard,Todorov} in THz cavities or the
cascade effect in quantum cascade lasers \cite{Faist} (QCL). Nevertheless,
until now THz sources operating in the spectral region around 1THz remain costly,
show limited quantum efficiency and operate at very low temperatures.

Recently, it was proposed that the efficiency of THz emission can be drastically increased by using the phenomenon of bosonic stimulation. This mechanism can be realized in excitonic and exciton-polaritonic\cite{KavokinBook} systems, where the elementary excitations participating in the THz transition have a bosonic nature ~\cite{Kavokin,Savenko,delValle,Cascade,DeLiberato}. The main benefits of THz radiators based on bosonic systems are their small size, relatively high quantum efficiency and the ability to operate at room-temperature.

A scheme which has attracted particular attention due to the possibility of realizing a vertical THz emitter is based on the possibility of a radiative transition between 2p and 1s exciton states \cite{Vertical}. As direct optical creation of the 2p exciton under single photon absorbtion is prohibited, it was proposed to use two-photon pumping of a $2p$ exciton state, as has been realized already in GaAs based quantum well
structures\cite{Cingolani, Schemla}. After creation, a $2p$ exciton can radiatively decay to
the $1s$ exciton state emitting a THz photon. The
inverse process (THz absorption by a lower polariton mode with excitation of
a $2p$ exciton) has been recently observed experimentally \cite{Tomaino}. The THz transition from the $2p$ state pumps the lowest energy exciton state, which can have macroscopic occupation and thus stimulate the THz emission. The situation can be further improved by replacing of the 2s exciton state by an exciton-polariton, the hybrid quasiparticle appearing in quantum microcavities \cite{KavokinBook}. In this case, THz emission eventually leads to the polariton lasing
effect, widely discussed in the literature and now routinely observed experimentally \cite{Christopoulos,Bajoni2008,Das,Sven}.

In this article we present the theory of a THz laser based on the 2p to 1s optical transition. We consider a quantum well placed inside a microcavity tuned close to the resonant frequency of two photon absorption. The presence of such a cavity should increase the intensity of the optical excitation of the dark 2p state and can even lead to multi-photon polaritonic effects as was shown recently in Ref. \onlinecite{Perv}. The structure is then placed in a THz cavity of larger size, which further increases the efficiency of THz emission by confining THz photons and thus leading to bosonic stimulation. We consider the case of coherent excitation at low temperatures. In this regime, one can suppose that effects of decoherence play a minor role, and the system should be described in terms of semiclassical laser equations corresponding to interacting coherent fields, and not in terms of Boltzmann equations, considered in Ref.\onlinecite{Vertical}, corresponding to the regime when decoherence is fast. We reveal some novel phenomena in the system provided by the nonlinearities of two photon absorption and THz emission, such as optical bistability and hysteresis of the THz signal.

\section{Geometry of the structure}
We consider the system formed by a quantum well supporting a 2p exciton state with energy $\hbar\omega_{2p}$ and a 1s exciton state with energy $\hbar\omega_s$, embedded within a planar microcavity supporting a photonic mode with energy $\hbar\omega_a=\hbar \omega_{2p}/2$ tuned into resonance with the two-photon optical transition to the 2p exciton state. The whole system is further embedded within a larger THz cavity with frequency $\hbar \omega_{THz}$ (Fig. 1). The system is excited resonantly by an external continuous wave laser beam of frequency $\omega$.

\begin{figure}[h!]
\centering
    \includegraphics[width=0.45\textwidth]{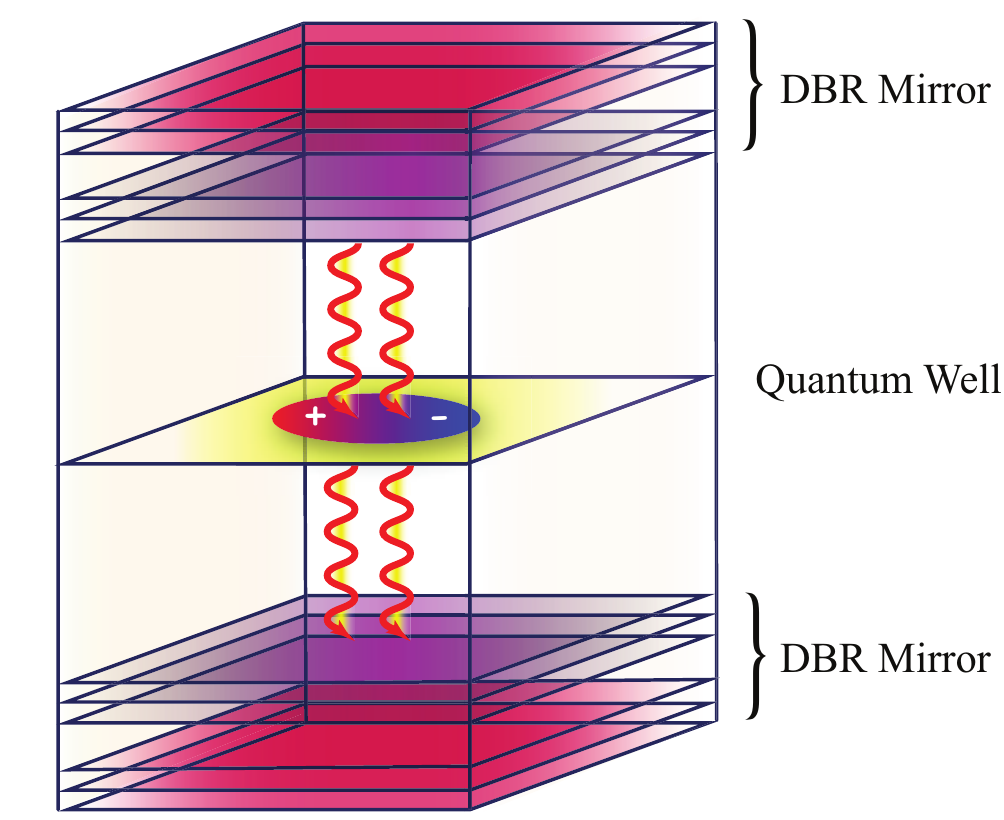}

\caption{Geometry of the structure. We consider a microcavity, which is made from two Bragg mirrors with a quantum well inside, where a $2p$ excitonic state can be excited by two photons with energy $\hbar\omega_{c}$ each.}
\end{figure}

The possible types of optical transitions in the system are shown in Fig. 2. The two photons with energy  $\hbar\omega_{c}$ each excite the 2p dark exciton level via virtual excitations of bright states indicated schematically by the horizontal dashed line in Fig. 2. It should be noted that direct single photon excitation of the 2p state violates the conservation of angular momentum and is thus forbidden. As 2p and 1s excitons have different parities, the radiative transition from the 2p exciton level to the 1s exciton level is possible and is accompanied by the emission of a THz photon with energy $\hbar\omega_{THz}$.

\begin{figure}[h!]
\centering
    \includegraphics[width=0.45\textwidth]{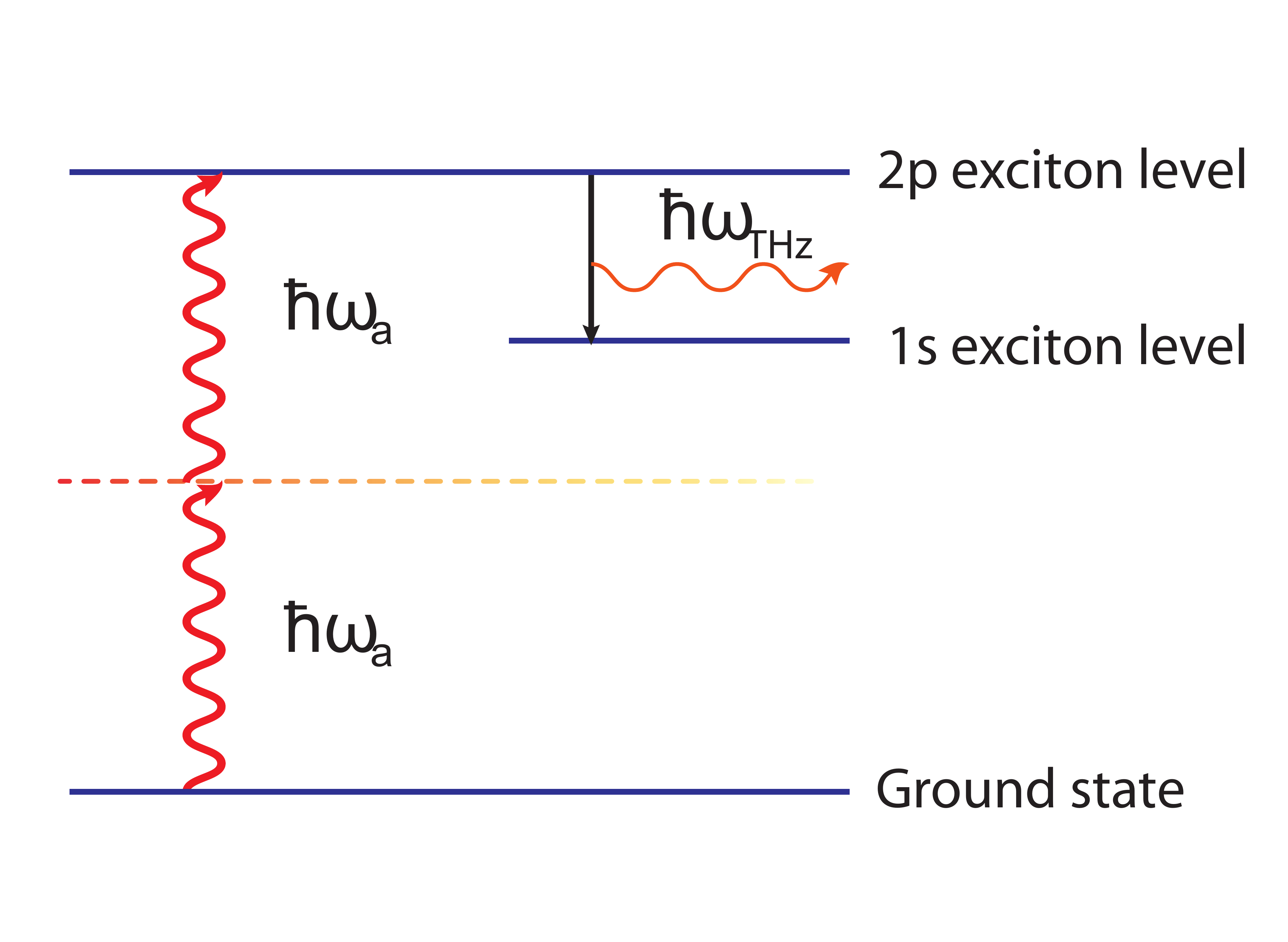}

\caption{The scheme of two-photon excitation of a 2p excitonic state and generation of THz radiation in a microcavity. The dark exciton state (2p state) can not be excited by a single photon due to optical selection rules. However, excitation is possible in the case of two-photon absorption with each photon having an energy $\hbar\omega_{c}$. Due to relaxation, 2p excitons decay into 1s exciton states, emitting THz photons with energy $\hbar\omega_{THz}$. Note that the illustration is not to scale and that in reality, $\hbar\omega_{c}\gg\hbar\omega_{THz}$.}
\end{figure}

The model Hamiltonian of this system within the rotating wave approximation can be written:
\begin{eqnarray}
\hat {\cal H}=\hbar\omega_p\hat{p}^+\hat{p}+\hbar\omega_a\hat{a}^+\hat{a}+\hbar\omega_s\hat{s}^+\hat{s}+\hbar\omega_{THz}\hat{c}^+\hat{c}+\nonumber\\
+g(\hat{p}\hat{a}^+\hat{a}^++\hat{p}^+\hat{a}\hat{a})+G\hat{p}^+\hat{s}\hat{c}+G^*\hat{p}\hat{s}^+\hat{c}^+,
\end{eqnarray}
where $\hat{p}$, $\hat{a}$, $\hat{s}$ and $\hat{c}$ are operators of 2p dark excitons, cavity photons, 1s bright excitons and terahertz photons, respectively, which satisfy bosonic commutation relations. The first four terms describe the energy of the free modes. The following term corresponds to the interaction between 2p excitons and optical photons. The last two terms describe THz emission associated with the 2p-1s transition. $g$ and $G$ are the coupling constants corresponding to two photon excitation and THz transition, respectively.

The two photon-exciton coupling constant $g$ was calculated previously using second-order perturbation theory~\cite{Perv}:

\begin{align}
g=\sqrt{\frac{S}{2}}{\left(-\frac{qA_0}{\mu}\right)}^2& \sum_{n}\frac{\frac{i\sqrt{E_g m_0^2}}{\sqrt{2m^*}}\Phi_n(0)\int d\vec{r}R_{21}(\vec r)\vec r R_{10}(\vec{r})}{2\hbar\omega - (E_g -E_n+\hbar\omega)}\times\nonumber\\
&\times\frac{im_0}{\hbar\sqrt{2}}(E_{2p}-E_{ns}),
\end{align}
where: $S$ is the quantization area of the sample; $m_0$, $m^*$ and $\mu$ are the free electron, effective and reduced exciton masses, respectively; $q$ is the elementary charge of the electron; $E_g$ is the band gap of the material; $\hbar\omega$ is the energy of the photon; $E_n$ are the eigenvalues of the 2D Hydrogen atom; $R_n(\vec r)$ and $\Phi_n(\vec r)$ are normalized radial and angular eigenfunctions of the 2D Hydrogen atom, respectively; and $A_0$ is the vector potential of the cavity photon mode. This last quantity can be written:

\begin{align}
A_0=&\sqrt{\frac{\hbar}{2\epsilon\epsilon_0\omega L S}},\nonumber
\end{align}
where $\epsilon_0$ and $\epsilon$ are vacuum and material permitivities; $L$ is the length of the planar microcavity supporting the photonic mode.

The constant of interaction between 2p and 1s exciton levels with THz lasing can be found from the matrix element of this transition in the dipole approximation ~\cite{Scully}, using the wave functions of the 2D Hydrogen atom ~\cite{Yang}:

\begin{align}
G^*=&\frac{1}{\sqrt{2}}\langle 1s|{\cal{\hat H}}|2p\rangle=\frac{1}{\sqrt{2}}\left(-\frac{qA'_0}{\mu}\right)\frac{im_0}{\hbar}\langle 1s|r|2p\rangle=\nonumber\\
=&-\frac{qA'_0}{\sqrt{2}\mu}\frac{27im_0\pi \epsilon\epsilon_0\hbar}{8\sqrt{6}\mu q^2}.
\end{align}
Here $A'_0$ is the vector potential of the THz photon mode, which is similar to $A_0$, but with a different length, $L'$ corresponding to the length of the THz cavity.
Here we consider a GaAs based structure with $L=800$ $nm$ and $L'=3.6$ $\mu m$.

It should be noted that this formalism can be applied generally to both schemes with 1s exciton states and polariton states. In the case of polariton states, the constants $g$ and $G$ should be multiplied by the corresponding Hopfield coefficients.

\section{Classical field approximation}

The dynamics of the system can be found by writing the Heisenberg equations of motion for the Hamiltonian in Eq.1:
\begin{align}
&i\hbar\frac{d\hat p}{dt}=\left[ \hat p,{\cal {\hat H}}\right]=\hbar\omega_p \hat p+g\hat a^2+G\hat s\hat c,\\
&i\hbar\frac{d\hat a}{dt}=\left[ \hat a,{\cal {\hat H}}\right]=\hbar\omega_a \hat a+2g\hat p\hat a^+ ,\\
&i\hbar\frac{d\hat s}{dt}=\left[ \hat s,{\cal {\hat H}}\right]=\hbar\omega_s \hat s+G^*\hat p \hat c^+,\\
&i\hbar\frac{d\hat c}{dt}=\left[ \hat c,{\cal {\hat H}}\right]=\hbar\omega_{THz} \hat c+G^*\hat p\hat s^+.
\end{align}
In the mean-field approximation one can take averages of the previous system of equations and switch to the mean values of the operators. The average of the product of the operators can be approximately replaced by the product of the mean values. Terms to describe the finite lifetime of the excitons and photons and external coherent pumping can be added to the equations phenomenologically. As a result of this approximation, the system of nonlinear differential equations of motion takes the following form:
\begin{align}
&i\hbar\frac{dp}{dt}=\hbar\omega_p p+ga^2+Gsc-i\gamma_p p,\\
&i\hbar\frac{da}{dt}=\hbar\omega_a a+2gpa^+ +P e^{-i\omega t}-i\gamma_a a,\\
&i\hbar\frac{ds}{dt}=\hbar\omega_s s+G^*p c^+-i\gamma_s s,\\
&i\hbar\frac{dc}{dt}=\hbar\omega_{THz} c+G^*ps^+-i\gamma_c c,
\end{align}

Here $p=\left<\hat p\right>$, $a=\left<\hat a\right>$, $s=\left<\hat s\right>$, and $c=\left<\hat c\right>$ are the mean values of the operators of 2p dark excitons, cavity photons, 1s bright excitons and terahertz photons, respectively. $\gamma_p$, $\gamma_a$, $\gamma_s$ and $\gamma_c$ are the decay rates for 2p dark excitons, cavity photons, 1s bright excitons and terahertz photons, respectively, which are related to the lifetimes of the modes,  $\tau_{p,a,s,c}=1/\gamma_{p,a,s,c}$. $P$ is the amplitude of the external coherent pump of the cavity photons with frequency $\omega$.

Making the substitution $p\rightarrow\tilde{p}(t)e^{-2i\omega t}$, $a\rightarrow\tilde{a}(t)e^{-i\omega t}$, $s\rightarrow\tilde {s}(t)e^{-i\omega_s t}$, $c\rightarrow\tilde{c}(t)e^{-i\omega_{THz} t}$ the system of equations can be rewritten as:

\begin{align}
&i\hbar\frac{d\tilde{p}}{dt}=(\hbar\Delta_p -i\gamma_p)\tilde{p}+g\tilde{a}^2+G \tilde{s}\tilde{c},\\
&i\hbar\frac{d\tilde{a}}{dt}=(\hbar\Delta_a -i\gamma_a)\tilde{a}+2g\tilde{p}\tilde{a}^*+P,\\
&i\hbar\frac{d\tilde{s}}{dt}=-i\gamma_s\tilde{s}+G^* \tilde{p}\tilde{c}^*,\\
&i\hbar\frac{d\tilde{c}}{dt}=-i\gamma_c\tilde{c}+G^*\tilde{p}\tilde{s}^*,
\end{align}
where $\Delta_p=\omega_p-2\omega, \Delta_a=\omega_a-\omega$.

\section{Stationary solutions}
In the stationary case $\frac{d\tilde{p}}{dt}=\frac{d\tilde {a}}{dt}=\frac{d\tilde{s}}{dt}=\frac{d\tilde {c}}{dt}=0$ and the system of differential equations transforms to a system of nonlinear algebraic equations.

This system of equations has two forms of solutions: the first one corresponds to the case without THz lasing, when the mean values of the operators $\hat s$ and $\hat c$ describing bright excitons and THz photons are equal to zero. The occupancy of THz mode is zero below threshold because the processes of spontaneous emission are neglected in our semiclassical theory. The solution without emission of THz photons can be written in terms of the real functions describing the number of photons, $N_a=|a|^2$ and excitons, $N_p=|p|^2$:
 \begin{align}
&N_a\left[1+c_1N_a+c_2N_a^2\right]=I_a,\\
&N_p=\frac{g^2N_a^2}{\gamma_p^2+\hbar^2\Delta_p^2},\notag
 \end{align}
 where
 \begin{align}
&c_1=\frac{4g^2(\gamma_a\gamma_p-\hbar^2\Delta_a\Delta_p)}{(\hbar^2\Delta_p^2+\gamma_p^2)(\hbar^2\Delta_a^2+\gamma_a^2)},\notag\\
&c_2=\frac{4g^4}{(\hbar^2\Delta_p^2+\gamma_p^2)(\hbar^2\Delta_a^2+\gamma_a^2)},\notag\\
&I_a=\frac{|P|^2}{\hbar^2\Delta_a^2+\gamma_a^2}.\notag
 \end{align}
This case corresponds to the one considered in our previous paper~\cite{Perv}. The solution demonstrates hysteresis of the mode populations as a function of the intensity of the pump. Differently from the case of conventional exciton-polariton bistability~\cite{Baas2004,Whittaker2005,Gippius}, the bistability comes from the nonlinearity of the two-photon absorption and not the exciton-exciton interaction although the latter can modify the bistable response curve.

The second solution corresponds to the situation when THz lasing occurs in the system. From Eqs.(14),(15) one has:
\begin{align}
&0=-i\gamma_s\tilde{s}+G^* \tilde{p}\tilde{c}^*,\\
&0=-i\gamma_c\tilde{c}+G^*\tilde{p}\tilde{s}^*,
\end{align}
which immediately gives for $\tilde s \neq 0$,  $\tilde c \neq 0$:
\begin{align}
&N_p=\frac{\gamma_s \gamma_c}{|G|^2}.
\end{align}
One sees that $N_p=const$, independent of the pumping intensity. Obviously, this solution has no physical meaning for small pumps, where indeed it always becomes unstable.

Equations (13-15) allow to obtain equations for the occupancies $N_a=|a|^2, N_s=|s|^2, N_c=|c|^2$:
\begin{align}
&(\hbar^2\Delta_a^2+\gamma_a^2+4g^2N_p)N_a-\notag\\
&-4N_p(\hbar^2\Delta_a\Delta_p-\gamma_a(\gamma_p+\frac{|G|^2N_c}{\gamma_s}))=|P|^2,\\
&N_c=\frac{\gamma_s}{|G|^2}(\sqrt{\frac{N_a^2|G|^2 g^2}{\gamma_s \gamma_c}-\hbar^2\Delta_p^2}-\gamma_p),\\
&N_s=\frac{\gamma_c N_c}{\gamma_s}=\frac{\gamma_c}{|G|^2}(\sqrt{\frac{N_a^2|G|^2 g^2}{\gamma_s \gamma_c}-\hbar^2\Delta_p^2}-\gamma_p).
\end{align}
The stability of the solutions can be tested by considering the behaviour of small fluctuations about the mean-field solutions~\cite{Whittaker2005}, of the form $\tilde{p}\mapsto\tilde{p}+u_pe^{-i\mu t}+v_p^*e^{i\mu^* t}$, with $u_p$ and $v_p$ complex fluctuation amplitudes and $\mu$ a complex frequency to be determined. Similar fluctuations are applied to $\tilde{a}$, $\tilde{s}$ and $\tilde{c}$. Substitution into Eqs. 12-15 and collection of terms oscillating as $e^{-i\mu t}$ and $e^{i\mu^* t}$ gives rise to a set of eight coupled equations. These can be solved for the eigenvalues $\mu$. If the imaginary part of $\mu$ is less than zero then the corresponding mean-field solution is stable, due to decay of the perturbation. On the other hand a positive imaginary part of $\mu$ corresponds to an unstable solution.

The stability of the solutions is governed by the intensity of the coherent pump. At some critical point of the pumping intensity the solution without THz lasing loses stability and the solution with THz lasing becomes stable. Furthermore, for some combination of the parameters one can see the interesting situation where both of the solutions are stable. In this case the behaviour of the system is defined by its history and the THz signal as a function of the pump reveals hysteresis behaviour as we show below.

The behaviour of the system is demonstrated in Figs. 3 and 4.

\begin{figure}[H]
\centering
    \includegraphics[width=0.5\textwidth]{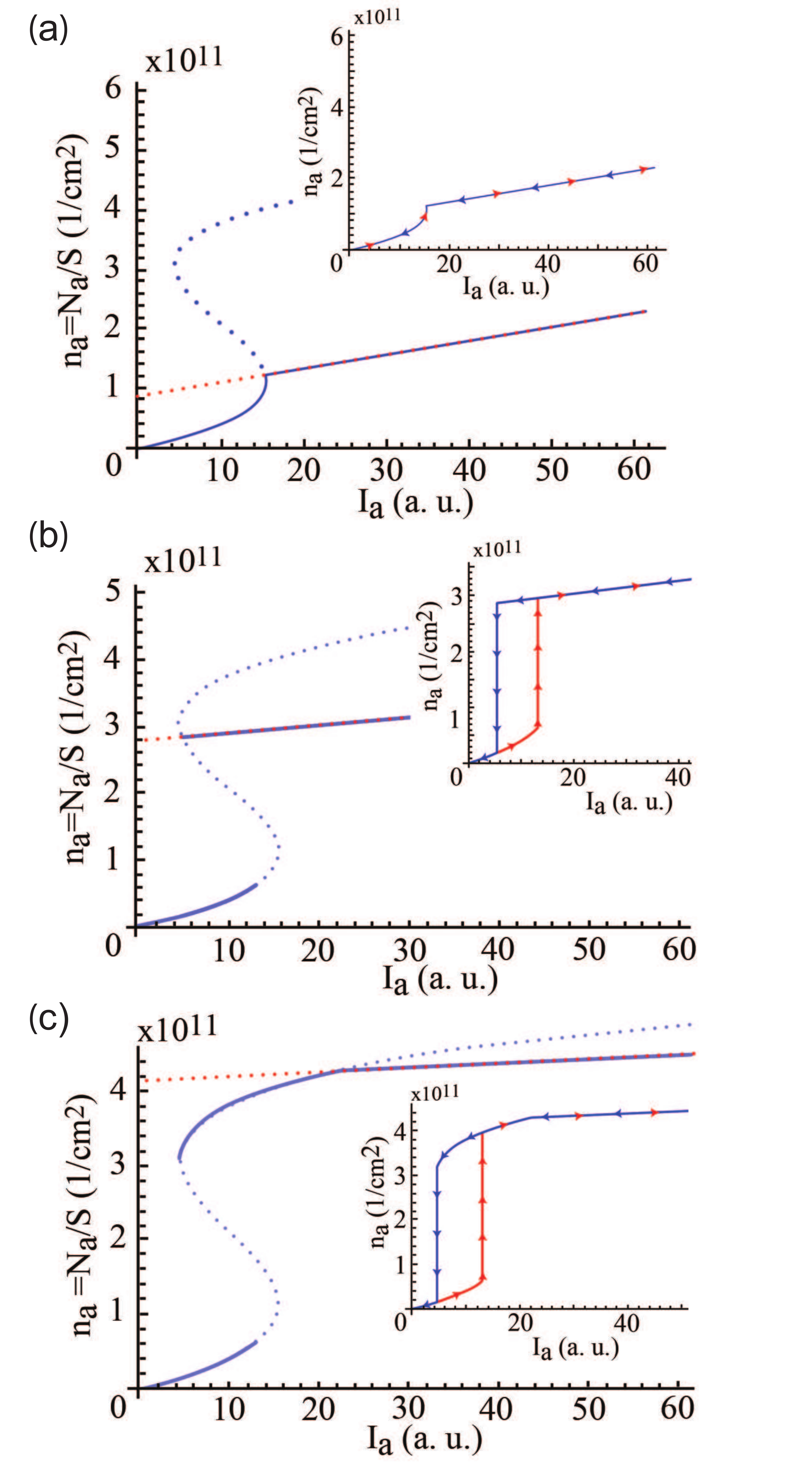}
\caption{ The dependence of the concentration of the cavity photons on the pumping intensity for the case when the quality of the THz resonator $\gamma_c=100\;\mu eV$(a), $\gamma_c=500\; \mu eV$(b), $\gamma_c=1100\;\mu eV$(c) . The two solutions for the system of equations (Eqs. (12-15)) are presented: solution without THz emission (blue dotted curve) and solution with THz emittion (red dotted line). The stable solutions of the problem, which describe the behavior of the system under coherent pump, are marked by the solid blue curves. The parameters are taken for the detuning $\hbar\Delta_a=0.25\;meV, \hbar\Delta_p=0.5\;meV$. In the insert, the stable solutions for the concentration of the cavity photons depending on the coherent pump intensity are presented. The direction of increasing and decreasing of the pump are indicated by arrows.}
\end{figure}
They show the occupancies of the optical cavity mode and THz mode, respectively as a function of the intensity of the resonant pump.
Keeping  $\gamma_p=0,01\;meV$, $\gamma_a=0,05\;meV$, $\gamma_s=0,05\;meV$ we studied how the behaviour of the system changes if the quality of the THz  resonator described by the parameter $\gamma_c$ changes. We consider three cases corresponding to the values $\gamma_c=100\;\mu eV$ (Figs. 3a, 4a); $\gamma_c=500\; \mu eV$ (Figs.3b, 4b); $\gamma_c=1100\; \mu eV$ (Figs. 3c, 4c).

Fig. 3a corresponds to a high quality THz cavity with $\gamma_c=100\; \mu eV$. The occupancy of the cavity photons as a function of the pump intensity is described by an S-shaped curve corresponding to the absence of the THz lasing (shown in blue) and straight line corresponding to the case of THz lasing (shown in red). For a given value of the parameter $\gamma_c$ the two lines intersect in the region of the lower stable branch of the S- shaped curve at some value of the pump $I_{th}$ corresponding to the threshold of THz lasing. If the pumping power is less than $I_{th}$, the solution with no THz lasing is stable, while the solution with lasing is unstable. The situation is inverted if the pumping power is more then $I_{th}$. There is no region of the pump where both solutions are stable, and thus the dependence of the occupancy of the THz mode on the intensity of the pump is a single valued function as shown in Fig. 4a.

The more interesting case corresponds to a medium- quality THz resonator with $\gamma_c=500\; \mu eV$, shown in Fig. 3b. The two solutions intersect in the unstable region of the S- shaped curve. Consequently, there is a region of pumping intensity where the solutions with and without THz lasing are stable. Therefore, the system demonstrates a hysteresis behaviour for both occupancies of the optical and THz modes, shown in Figs. 3b and 4b, respectively.

Fig. 3c corresponds to the case of a low-quality THz resonator with $\gamma_c=1100\; \mu eV$. In this case the curves corresponding to the situation without and with THz lasing intersect in the region of the upper stable branch of the S-shaped curve. The dependence of the occupancy of the cavity photons on the intensity of the pump thus demonstrates a hysteresis behavior. The hysteresis loop, however, lies entirely in the region when THz lasing does not occur. The occupancy of the THz mode still remains a single-valued function of the pump as shown in Fig. 4c.

\begin{figure}[h!]
\centering
    \includegraphics[width=0.5\textwidth]{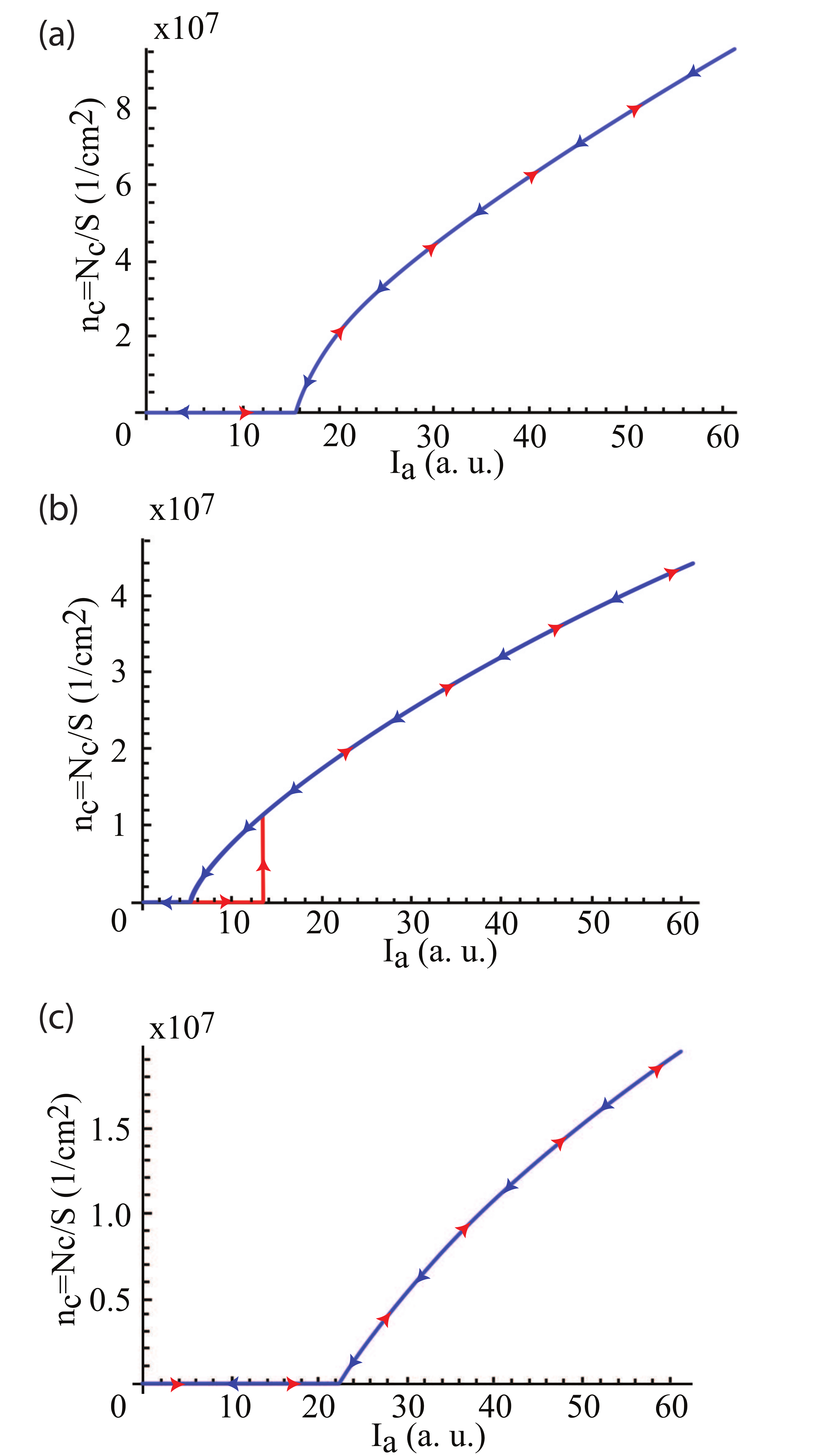}
\caption{ The dependence of the concentration of THz photons on the pumping intensity for the case when the decay rate of the THz resonator $\gamma_c=100\; \mu eV$(a), $\gamma_c=500\; \mu eV$(b), $\gamma_c=1100\; \mu eV$(c). The direction of increasing and decreasing of the pump is indicated by arrows.}
\end{figure}

\section{Conclusions}

In conclusion, we developed theoretically a model of a THz laser based on the radiative transition between 2p and 1s excitonic states. We have shown that nonlinearities provided by two-photon absorption and THz photon emission strongly affect the behavior of the system, and can lead to the onset of bistable behaviour of the THz signal as a function of the external pump. An important parameter which governs the regime of the operation of the device is the quality factor of the THz cavity.

We thank Dr. Ivan Savenko and Oleksandr Kyriienko for discussions. The work was supported by FP7 IRSES project POLATER and FP7 ITN NOTEDEV.



\begin{thebibliography}{99}

\bibitem{Dragoman}D. Dragoman and M. Dragoman, Prog. Quantum Electron. \textbf{28}, 1 (2004).

\bibitem{Mueller} E. R. Mueller, The Ind. Phys. \textbf{9}, 27 (2003).

\bibitem{Hu} Q. Hu, B. S. Williams, S. Kumar, H. Callebaut, S. Kohen, and J. L. Reno, Semicond. Sci. Technol. \textbf{20}, S228 (2005).

\bibitem{Portnoi}M. E. Portnoi, O. V. Kibis, M. R. da Costa, Superlatt. Microstruct. \textbf{43}, 399 (2008).

\bibitem{Wright} A. R. Wright, J. C. Cau, and C. Zhang, Phys. Rev. B. \textbf{74}, 165328 (2004).

\bibitem{Purcell} E.M. Purcell, \emph{Phys. Rev.} \textbf{69}, 681 (1946)

\bibitem{Gerard} J.-M. Gerard and B. Gayral, \emph{Journ. Lightwave Technol.}
\textbf{17}, 2089 (1999)

\bibitem{Todorov} Y. Todorov \emph{et al}, \emph{Phys.
Rev. Lett.} \textbf{99}, 223603 (2007).

\bibitem{Faist} R.F. Kazarinov, and R.A. Suris, \emph{Sov. Phys.
Semiconductors} \textbf{5}, 707 (1971); J. Faist \emph{et al},
\emph{Science}, \textbf{264,}
553 (1994); E. Normand \emph{et al}, \emph{Laser Focus World}%
, \textbf{43,} 90 (2007).

\bibitem{Kavokin} K. V. Kavokin, M. A. Kaliteevski, R. A. Abram, A. V. Kavokin, S. Sharkova, and I. A. Shelykh, Appl. Phys. Lett. \textbf{97}, 201111 (2010).

\bibitem{Savenko} I. G. Savenko, I. A. Shelykh, and M. A. Kaliteevski, Phys. Rev. Lett. \textbf{107}, 027401 (2011).

\bibitem{delValle} E. del Valle and A. Kavokin, Phys. Rev. B, {\bf 83}, 193303 (2011)
\bibitem{Cascade} T. C. H. Liew, M. M. Glazov, K. V. Kavokin, I. A. Shelykh, M. A. Kaliteevski, A. V. Kavokin, Phys. Rev. Lett. 110, 047402 (2013)

\bibitem{DeLiberato} S. De Liberato, C. Ciuti, and C. C. Phillips, Phys. Rev. B, {\bf 87}, 241304(R) (2013).

\bibitem{Vertical} A. V. Kavokin, I. A. Shelykh, T. Taylor, and M. M. Glazov, Phys. Rev. Lett. \textbf{108}, 197401 (2012).

\bibitem{Cingolani} I.M. Catalano \textit{et al}, Phys. Rev B \textbf{40},
1312 (1989).

\bibitem{Schemla} R. A. Kaindl, D. Hagele, M. A. Carnahan, and D. S. Chemla,
Phys. Rev. B \textbf{79}, 045320 (2009).

\bibitem{KavokinBook} A.V. Kavokin, J.J. Baumberg, G. Malpuech and F.P.
Laussy, Microcavities, Oxford University Press, Oxford (2007).

\bibitem{Tomaino} J. L. Tomaino, A. D. Jameson, Yun-Shik Lee, G. Khitrova, H. M. Gibbs, A. C. Klettke, M. Kira, and S. W. Koch, Phys. Rev. Lett. \textbf{108}, 267402 (2012).

\bibitem{Christopoulos} see e.g. S. Christopolous \textit{et al}, Phys. Rev.
Lett.\textbf{\ 98}, 126405 (2007)

\bibitem{Bajoni2008}
D Bajoni, \textit{et al.}, Phys. Rev. Lett., {\bf 100}, 047401 (2008).

\bibitem{Das} A. Das \textit{et al, }Phys. Rev. Lett.%
\textit{\ }\textbf{107}, 066405 (2011).

\bibitem{Sven} C. Schneider, A. Rahimi-Iman, Na Young Kim, J. Fischer, I. G. Savenko, M. Amthor, M. Lermer, A. Wolf, L. Worschech, V.D. Kulakovskii, I.A. Shelykh, M. Kamp, S. Reitzenstein, A. Forchel, Y. Yamamoto and S. Hofling, Nature \textbf{497}, 348 (2013)

\bibitem{Perv} A. A. Pervishko, T. C. H. Liew, V. M. Kovalev, I. G.Savenko, and I. A. Shelykh, Opt. Express \textbf{21}(13), 15183-15194 (2013).

\bibitem{Scully} Marlan O. Scully, M. Suhail Zubairy, Quantum Optics, Cambridge Univ. Press, 1997.

\bibitem{Yang} X. L. Yang, S. H. Guo, F. T. Chan, K. W. Wong, W. Y. Ching, Phys. Rev. A \textbf{43}, 3, (1991).

\bibitem{Baas2004}
A. Baas, J. P. Karr, H. Eleuch, and E. Giacobino, Phys. Rev. A {\bf 69}, 23809 (2004).

\bibitem{Whittaker2005}
D. M. Whittaker, Phys. Rev. B {\bf 71}, 115301 (2005).

\bibitem{Gippius} N. A. Gippius, I. A. Shelykh, D. D. Solnyshkov, S. S. Gavrilov, Yuri G. Rubo, A. V. Kavokin, S. G. Tikhodeev, G. Malpuech, Phys. Rev. Lett. \textbf{98}, 23, 2007.


\end{thebibliography}
\end{document}